\RequirePackage{lineno}
\documentclass[twocolumn,showpacs,superscriptaddress,amsmath,amssymb,nofootinbib]{revtex4-2}
\usepackage[x11names]{xcolor}
\usepackage{caption}
\usepackage[colorlinks, linkcolor=blue, citecolor=OliveGreen, urlcolor=black]{hyperref}
\usepackage{orcidlink}

\allowdisplaybreaks[4]
\usepackage{graphicx}
\usepackage{subfigure}
\usepackage{subcaption}
\usepackage{epsfig}
\usepackage{overpic}
\usepackage{dcolumn}
\usepackage{ulem}
\usepackage{bm}
\usepackage{color}
\usepackage{lineno}
\usepackage{xspace}
\usepackage{multirow}
\usepackage{epstopdf}
\usepackage{xcolor}
\usepackage{soul}
\usepackage{verbatim}
\usepackage{enumitem}
\usepackage{todonotes}
\usepackage{slashed}
\usepackage{cancel}
\usepackage{array}
\usepackage{amsmath}
\usepackage[toc,page]{appendix}

\definecolor{OliveGreen}{rgb}{0.4, 0.8, 0.1}

\usepackage{diagbox}

\newcommand{\moe}{\affiliation{Key Laboratory of Atomic and Subatomic Structure and 
Quantum Control (MOE), Guangdong-Hong Kong Joint Laboratory of Quantum Matter, Guangzhou 510006, China}}

\newcommand{\iqm}{\affiliation{State Key Laboratory of Nuclear Physics and 
Technology, Institute of Quantum Matter, South China Normal 
University, Guangzhou 510006, China}}

\newcommand{\gbrce}{\affiliation{Guangdong Basic Research Center of Excellence for 
Structure and Fundamental Interactions of Matter, Guangdong Provincial Key Laboratory of Nuclear Science, Guangzhou 510006, China}}

\newcommand{\scnt}{\affiliation{Southern Center for Nuclear-Science Theory (SCNT), Institute of Modern Physics, Chinese Academy of Sciences, Huizhou 516000, Guangdong Province, China}}

\newcommand{\fudana}{\affiliation{Key Laboratory of Nuclear Physics and Ion-beam Application (MOE), Institute of Modern Physics, Fudan University, Shanghai 200433, China}}

\newcommand{\fudanb}{\affiliation{Shanghai Research Center for Theoretical Nuclear Physics, NSFC and Fudan University, Shanghai 200438, China}}

\begin{document}
\include{def-com}

\title{\boldmath Thermal width shift of $\Delta^{++}$ in a pion gas}

\author {\mbox{Ying Zhang\orcidlink{0009-0008-9322-1625}}}
\email{yingzhang@m.scnu.edu.cn}
\iqm
\moe
\gbrce

\author {\mbox{Peng-Yu Niu\orcidlink{0000-0001-8455-9570}}}
\email{niupy@m.scnu.edu.cn}
\iqm
\gbrce

\author {\mbox{Xin-yue Hu}\orcidlink{0009-0005-2255-4543}}
\email{xinyue\_hu@m.scnu.edu.cn}
\iqm
\moe
\gbrce

\author {\mbox{Kai-Jia Sun}\orcidlink{https://orcid.org/0009-0009-5531-9274}}
\email{kjsun@fudan.edu.cn}
\fudana
\fudanb

\author {\mbox{Qian Wang\orcidlink{0000-0002-2447-111X}}}
\email{qianwang@m.scnu.edu.cn}
\iqm
\gbrce
\scnt

\date{\today}

\begin{abstract}
We compute the thermal width shift of the $\Delta^{++}$ resonance induced by a pion gas within a nonrelativistic effective field theory framework. The $\Delta^{++}$ self-energy is evaluated from pion-forward-scattering diagrams with intermediate proton and $\Delta$ states, weighted by the thermal pion distribution. Analytical expressions for the imaginary part of the self-energy yield the temperature-dependent width correction $\delta\Gamma(T)$. The width increases with temperature, reaching approximately $6$~MeV at $T \approx 160$~MeV. When the temperature-dependent $\Delta$ and nucleon masses from an NJL-model chiral restoration scenario are incorporated, the width shift becomes non-monotonic, peaking near $T \approx 140$~MeV---a consequence of the competition between collisional broadening and the shrinking $\Delta \to N\pi$ phase space as the $N$--$\Delta$ mass gap closes. Applying our formalism to STAR data for $\Delta^{++}$ in d+Au collisions at $\sqrt{s_{NN}} = 200$~GeV, we extract a temperature $T \approx 300$~MeV at $p_t = 900$~MeV, consistent with the experimental extraction within errors. This value significantly exceeds the hadronic-phase temperature, explicitly demonstrating that the observed $\Delta^{++}$ width shift is not solely of pion-gas origin---genuine hot-medium and collective-flow contributions must be substantial. 
\end{abstract}
\maketitle
\section{Introduction}
\label{Intro}

Quantum chromodynamics (QCD) is a fundamental theory that describes the strong interactions occurring between quarks and gluons. It hints that the hadronic matter would be deconfined and transit to the quark-gluon plasma (QGP) phase in the environment at high temperatures and/or high densities~\cite{Cheng:2006qk, Bernard:2006nj, Karsch:2007vw}. Before the period of deconfinement, resonances are under a special condition where their physical properties, such as masses~\cite{Montana:2020lfi, Gu:2018swy}, widths~\cite{Montana:2020lfi, Fuchs:2004fh, Cleven:2017fun}, and even their spectral shapes~\cite{Brown:1991kk, Rapp:2003ar, Shuryak:2002kd}, will be modified by the in-medium effects related to the high temperature and/or high density. After chemical freeze-out, these properties can also be modified by surrounding hadrons. To isolate the pure effect arising from the QGP, we need to disentangle contributions from the hadronic environment. The widths of resonances will be modified by the in-medium effects, which has drawn considerable research interests and yielded a wealth of findings: the width of light mesons is increasing with increasing nuclear density in cold nuclear matter produced by $p$+A collisions~\cite{KEK-PS-E325:2005wbm} or in hot and dense hadronic matter produced by heavy ion collisions~\cite{Ruppert:2005id,Rapp:2010sj}; the width of heavy quarkonium moving in high-temperature QCD plasmas is closely related to the diffusion coefficient and dispersive
counterpart of heavy quarks~\cite{Hong:2022ksb}; the width broadening of $K^*$, $\phi$, $\Delta$ and $\Lambda$ in quark-gluon phase at RHIC can be attributed to chiral symmetry restoration~\cite{Markert:2008jc}.

Studies on the physical properties of resonances in medium with high temperature and/or high density can be accomplished by relativistic heavy ion collisions (HICs)~\cite{Schaffner-Bielich:1999cux, Rapp:1999ej}, where a transition from a deconfined QGP to a system of hadrons in a thermal medium
occurs for high-energy collision events. HIC experiments are now undergoing a promising generation of operational or under construction, such as LHC, EIC, and FAIR. It is noted that the Relativistic Heavy-Ion Collider (RHIC) has provided a variety of collision systems at different beam energies, including collisions of Cu+Cu, Au+Au, $d$+Au and $p$+$p$ at $\sqrt{s_{NN}}=200$ GeV~\cite{STAR:2004bgh, PHENIX:2010bqp, STAR:2003pjh, STAR:2003wqp, STAR:2002svs}. Using the data from RHIC, the experimentalists have successfully analyzed the production of some resonances, including both the light flavors and the heavy ones. For references, we recommend Refs.~\cite{Gopal:2022zgm, Das:2026qnu, STAR:2024znc,Tang:2020ame, Knospe:2026kiq}. 

The $\Delta$(1232) resonance is the lightest member of the baryon decuplet and the dominant degree of freedom governing pion dynamics in hadronic matter at intermediate energies. Its vacuum properties are well established: a Breit--Wigner mass of $M_\Delta \approx 1232$~MeV and a decay width of $\Gamma_\Delta \approx 117$~MeV, overwhelmingly into the $\Delta \to N\pi$ channel. However, when the $\Delta$ resonance is immersed in a pion-rich environment---as realized in heavy-ion collisions, in the late stages of neutron-star mergers, or in the hadronic phase of the QCD fireball---its spectral properties undergo significant modifications. Understanding how and why the $\Delta$ width shifts in such an environment is not merely a question of a few tens of MeV in a resonance parameter; it carries far-reaching consequences that connect microscopic hadronic dynamics to the macroscopic signals of the QCD phase structure. This article is devoted to a systematic investigation of the in-medium width of the $\Delta$ resonance in a hot and dense pion gas. Here, we outline why this problem is important, structuring the discussion around four interconnected themes.

\medskip\noindent
\textbf{1. Chemical equilibration rate and detailed balance.}
In the hadronic stage of a relativistic HIC, the most efficient channel for establishing chemical equilibrium between pions and nucleons is the resonant reaction $\Delta \leftrightarrow N + \pi$. The rate at which this equilibrium is approached is directly controlled by the in-medium width of the $\Delta$,
\begin{equation}
\Gamma_{\Delta \to N\pi}^{\text{med}} \; f_\Delta \;=\; \langle \sigma_{N\pi \to \Delta} \, v_{\rm rel} \rangle \; f_N f_\pi \;,
\label{eq:equilibrium}
\end{equation}
where $f_i$ denote the phase-space distributions. In state-of-the-art hadronic transport models such as UrQMD~\cite{Bass:1998ca}, SMASH~\cite{SMASH:2016zqf}, and GiBUU~\cite{Buss:2011mx}, the detailed-balance condition is enforced at each collision vertex. If the vacuum Breit--Wigner width were employed without medium corrections, the inverse $N\pi \to \Delta$ cross section would systematically deviate from experimental measurements~\cite{Bleicher:1999xi}. As demonstrated explicitly in the SMASH validation study~\cite{SMASH:2016zqf}, the self-consistency of detailed balance in a box calculation is restored only when the resonance width is allowed to depend on the local medium conditions. Consequently, the in-medium $\Delta$ width is not an adjustable refinement but a prerequisite for any transport model that aims to correctly predict pion and nucleon yields across the beam-energy range $E_{\rm kin} = 0.4$--$2A$~GeV~\cite{Bratkovskaya:2000qy}.

\medskip\noindent
\textbf{2. Energy dependence of pion production.}
The medium-modified $\Delta$ width directly imprints itself on the experimentally measured pion production cross sections as a function of beam energy. In the subthreshold regime (where the collision energy per nucleon is below the free $NN \to NN\pi$ threshold), the effective $\Delta$ width must be increased by $50$--$80$~MeV beyond its vacuum value in order to reproduce the pion yields measured by FOPI and HADES~\cite{Cassing:1990dr, Cozma:2014yna}. In the resonance region ($1$--$2A$~GeV), the $\pi^-/\pi^+$ ratio measured in Au+Au collisions by HADES~\cite{Agakishiev:2010zw} shows systematic deviations from transport-model predictions that can be traced back to the isospin-dependent threshold corrections of the four $\Delta$ charge states ($\Delta^{++}$, $\Delta^+$, $\Delta^0$, $\Delta^-$) in the medium~\cite{STAR:2013pwb}. These observations demonstrate that the in-medium width shift is not a simple energy-independent scaling factor, but a complex function of density, temperature, isospin asymmetry, and the relative momentum of the $\Delta$ with respect to the surrounding pions~\cite{Buss:2006vh}. Any attempt to approximate this shift by a constant additional width inevitably fails to describe the full set of excitation functions.

\medskip\noindent
\textbf{3. Chiral symmetry restoration as a driving mechanism.}
At a more fundamental level, the in-medium modification of the $\Delta$ width is linked to the partial restoration of chiral symmetry in hot and dense hadronic matter. As the chiral condensate $\langle\bar{q}q\rangle$ melts with increasing temperature and/or density, the mass splitting between the nucleon and the $\Delta$ is expected to shrink~\cite{Cohen:1989qe, Brown:2001nh}. A reduced $N$--$\Delta$ mass gap implies a smaller phase space for the dominant $\Delta \to N\pi$ decay, which would tend to \textit{narrow} the width---counteracting the purely kinetic broadening mechanisms discussed above~\cite{Jido:2002yb}. The net width change near the chiral crossover is therefore the result of a delicate cancellation between the narrowing driven by chiral mass degeneracy and the broadening caused by enhanced multi-body collisions as the pion density grows~\cite{Leupold:2009kz}. Recent QCD sum-rule analyses suggest that this cancellation renders the $\Delta$ width a comparatively mild chiral probe---unlike the $\rho$ meson mass, which drops dramatically, the $\Delta$ width may remain relatively flat across the phase transition~\cite{Hohler:2013eba}. Precisely for this reason, a precision measurement of the $\Delta$ width as a function of collision centrality (i.e., as a function of the attained temperature and density) could provide a clean window into chiral dynamics that is less contaminated by trivial thermal effects.

Luckily, the STAR Collaboration presents the first measurement of the mass and width of $\Delta^{++}$ in $d$+Au collisions at $\sqrt{s_{NN}}=200$ GeV at RHIC~\cite{STAR:2008twt}, which shows a visible shift of mass and width in the distribution of transverse momentum $p_t$. The corresponding high temperature parameter $T$ extracted in the experiment~\cite{STAR:2008twt} suggests the probability of in-medium effects on the shift of its mass and width. Besides, Braaten \textit{et al}~\cite{Braaten:2023vgs} theoretically obtain the mass and width shifts of charmed mesons as a function of temperature $T$ by considering the pion gas modification to their self energies. Their results show a visible correction to the masses and widths of charmed mesons from the in-medium effects. 
Therefore, we would like to apply the same technique to explore the width shift of $\Delta$ in pion gas. The mass shift of $\Delta$ and nucleon are borrowed from Ref.~\cite{Torres-Rincon:2015rma}. Assuming the \(\Delta^{++}\) resides within a pion gas, we evaluate the thermal width shift of the \(\Delta^{++}\). In this analysis, medium modifications to the \(\Delta^{++}\) self-energy originating from \(\Delta^{++}\)–pion interactions are taken into account. 

The paper is organized as follow: in Sec.~\ref{Formalism}, we present the propagator corrections of $\Delta^{++}$ from its self energy, the calculation of its self energy in the pion gas, and the extraction of the corresponding width shift. The theoretical results and discussions follow as Sec.~\ref{Results and Analysis}. The summary is at the end as Sec.~\ref{summary}. Some details of the calculations are presented in the Appendixes.

\section{Formalism}\label{Formalism}
In this section, we present the corrections from the pion gas to the $\Delta^{++}$ width. More specifically, we calculate its self energy in the pion gas and extract the corresponding thermal width shift. 
\subsection{Propagator Corrections}
$\Delta^{++}$ is a baryon decuplet with quantum number  $J^{P}=\dfrac{3}{2}^+$, so its free Feynman propagator is given by~\cite{Yao:2016vbz, Hacker:2005fh} 
\begin{align}
    iS_{0,ij}^{\mu\nu}(p)=\xi^{3/2}_{ij}iS_0^{\mu\nu}(p),
\end{align}
where
\begin{align}
    iS_{0}^{\mu\nu}(p)
    =&-\frac{i(\slashed{p}+M_{\Delta})}{p^2-M_{\Delta}^2+i\epsilon}\Big[g^{\mu\nu}-\frac{1}{d-1}\gamma^\mu\gamma^\nu+\frac{1}{(d-1)M_{\Delta}}\notag\\
    &~\times(p^\mu\gamma^\nu-\gamma^\mu p^\nu)-\frac{d-2}{(d-1)M_{\Delta}^2}p^\mu p^\nu\Big],
    \label{eq:S0_complete}
\end{align}
with $p$ and $M_{\Delta}$ being the four momentum and mass of $\Delta^{++}$, respectively. $\xi_{ij}^{3/2}$ is the isospin-$\dfrac{3}{2}$ projection operator. $d$ denotes the space-time dimension and is set to 4 in our calculation. $\epsilon$ is a positive infinitesimal number that ensures the analyticity of Eq.~\eqref{eq:S0_complete} when $p=M_\Delta$. Since the relativistic calculation is full of complexity, $\Delta^{++}$ is treated as a nonrelativistic particle, and its nonrelativistic propagator becomes as
\begin{align}
    iS_{0}^{ij}(E,\bm{p})
    =~\frac{-iI}{E-\varepsilon-\bm p^2/(2M_{\Delta})+i\epsilon}\left[-\delta^{ij}+\frac{1}{3}\sigma^i\sigma^j\right],
    \label{propagator_Delta_NR}
\end{align}
where 
\begin{align}
    I=\left(\begin{array}{cc}
        1_{2\times2} &0  \\
         0& 0
    \end{array}\right).\label{eq:I}
\end{align}
Here, $\varepsilon=M_\Delta-M$ is the rest energy with $M$ the isospin-averaged mass of $\Delta$ particles, and then it is comparable to the isospin splitting. $E$ is energy relative to $M$.

\begin{figure}[h]
    \centering  \includegraphics[width=1.1\linewidth]{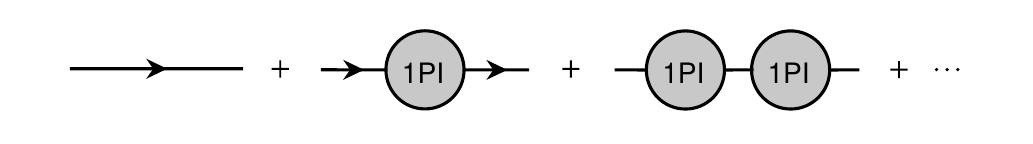}
    \caption{Correction from one-particle-irreducible (1PI) diagrams to $\Delta^{++}$ propagator.}
    \label{fig:1PI}
\end{figure}

Interactions of the \(\Delta^{++}\) with other particles such as protons and pions induce corrections to the pole position of the \(\Delta^{++}\) in its free Feynman propagator, which can be encoded by the self-energy \(\Sigma(E,\bm{p})\). 
Here $i\Sigma^{ij}(E,\bm{p})$ originates from the sum of one-particle-irreducible (1PI) diagrams contributing to the $\Delta^{++}$ propagator. Therefore, as shown by Fig.~\ref{fig:1PI}, the dressed $\Delta^{++}$ propagator can now be written as 
\begin{align}
    iS^{ij}(E,\bm{p})=&~iS_0^{ij}(E,\bm{p})+iS_0^{ik}(E,\bm{p})i\Sigma_{kl}(E,\bm{p})\notag\\
    &\times iS_0^{lj}(E,\bm{p})+\ldots+,\label{eq:S_ij1}
\end{align}
which can be obtained by solving the equation \cite{Hacker:2005fh}
\begin{align}
    iS^{ij}(E,\bm{p})=iS_0^{ij}(E,\bm{p})-iS^{ik}(E,\bm{p})\Sigma_{kl}(E,\bm{p})S_0^{lj}(E,\bm{p}).
\end{align}
The dressed nonrelativistic $\Delta^{++}$ propagator becomes 
\begin{align}
  iS^{ij}(E,\bm{p})=&~\frac{iS_0^{ij}(E,\bm{p})\Big(E-\varepsilon-\bm p^2/(2M_{\Delta})+i\epsilon\Big)}{E-\varepsilon-\bm p^2/(2M_{\Delta})-\Sigma(E,\bm{p})+i\epsilon}\notag\\
  &+\text{pole-free terms}.  
  \label{eq:S_ij2}
\end{align}
Note that $S_0^{ij}(E,\bm{p})\Big(E-\varepsilon-\bm p^2/(2M_{\Delta})+i\epsilon\Big)$ has no pole and $\Sigma(E,\bm{p})$ modifies the pole position of $\Delta^{++}$ propagator.

To make a clearer clarification, the effects of the self energy $\Sigma(E,\bm{p})$ are further indicated by a shift $\delta\varepsilon$ in the rest energy of $\Delta^{++}$, a correction $\delta Z$ to the residue of the pole in its propagator, and a multiplicative factor $1+\zeta$ to the inverse kinetic mass $\dfrac{1}{M_\Delta}$~\cite{Braaten:2023vgs}. With the above definition, the behavior of the dressed $\Delta^{++}$ propagator near its pole is 
\begin{align}
   iS^{ij}(E,\bm{p})=&~\frac{iS_0^{ij}(E,\bm{p})\Big(E-\varepsilon-\bm p^2/(2M_{\Delta})+i\epsilon\Big)(1+\delta Z)}{E-(\varepsilon+\delta\varepsilon)-(1+\zeta)\bm p^2/(2M_{\Delta})+\ldots}\notag\\
   &+\text{pole-free terms},\label{eq:S_ij3}
\end{align}
where ``..." represents the additional corrections to the denominator, which are the higher order contributions with an expansion in powers of $\bm{p}^2$, i.e., with powers starting from order $\bm{p}^4$. The factor $1+\delta Z$ in the numerator can be obtained by requiring the coefficient of $E$ in the denominator to be 1. Obviously, $\delta\varepsilon$ is related to the mass correction of $\Delta^{++}$ from the self energy and can be complex, indicating that its imaginary part can provide modifications to the width of $\Delta^{++}$ from the environment through the self energy. Therefore, to investigate the width shift of $\Delta^{++}$ in $d+$Au collisions, we will concentrate on the calculation of its corresponding $\text{Im}[\delta\varepsilon]$.

\subsection{Thermal Width Shift of $\Delta^{++}$ in the Pion Gas}
In this section, the framework of the width shift of $\Delta^{++}$ baryon in the pion gas is presented. 

\subsubsection{Thermal averages in the pion gas}
A region of QGP can be produced in the $d$+Au collisions at $\sqrt{s_{NN}}=200$ GeV at RHIC in the early stages, and then it expands and cools in thermal equilibrium. When it reaches the temperature of hadronization, it transits to a hadron resonance gas and continues to expand and cool in thermal equilibrium until its kinetic freeze-out. At temperatures $T$ near the temperature of kinetic freeze out $T_{\text{k}}$, the hadron resonance gas is dominated by the pions. The abundance of kaons is less than that of pions by a factor of 5, and the other hadrons are even less abundant~\cite{Braaten:2023vgs}, so the hadron gas can be approximated by a pion gas. The four momentum distribution of the pions is a Bose-Einstein distribution, i.e., $\bm{f}_\pi(\omega_q)=1/(e^{\omega_q/T}-1)$, where $\omega_q=\sqrt{m_\pi^2+\bm{q}^2}$ with $m_\pi$ the isospin-averaged mass of the pions. Thus, the number density of the pions in thermal equilibrium at temperature $T$ is 
\begin{align}
    \bm{n}_\pi^{\text{eq}}=\int \dfrac{\text{d}^3q}{(2\pi)^3}1/(e^{\omega_q/T}-1).
\end{align}
After kinetic freeze-out, the hadron gas continues to expand at a fixed $T=T_{\text{k}}$ with a decreasing number density due to a free-streaming of hadrons and is no longer in thermal equilibrium. Therefore, the pion momentum distribution can be described by~\cite{Braaten:2023vgs} 
\begin{align}
    \bm{f}_\pi(\omega_q)=\dfrac{\bm{n}_\pi}{\bm{n}_\pi^{\text{eq}}}\dfrac{1}{e^{\omega_q/T}-1}
\end{align}
for two scenarios: the expanding and cooling pion gas prior to kinetic freeze out, as well as the expanding pion gas after kinetic freeze out. 

The thermal average of a physical quantity $F(\bm{q})$ over the pion momentum is defined as~\cite{Braaten:2023vgs}
\begin{align}
    \langle F(\bm{q})\rangle=\int \dfrac{\text{d}^3q}{(2\pi)^3}\bm{f}_\pi(\omega_q)F(\bm{q})\bigg/\int \dfrac{\text{d}^3q}{(2\pi)^3}\bm{f}_\pi(\omega_q),
\end{align}
where the angular bracket represents the average over the pion momentum distribution. It is noted that the thermal average is only dependent on temperature $T$, but independent of the pion number density $\bm{n}_\pi$. Here we list all the thermal averages used in our calculations afterwards:
\begin{align}
    \mathcal{I}(T)&=\bigg\langle\dfrac{\bm{q}^2\omega_q}{\omega_{q}^2-\Delta_{p\Delta}^2+i\epsilon}\bigg\rangle,\label{def:I} \\    \mathcal{F}(T)&=\bigg\langle\dfrac{\bm{q}^2}{\omega_q}\dfrac{1}{\omega_q^2-\Delta_{p\Delta}^2+i\epsilon}\bigg\rangle,\label{def:F}\\
    \mathcal{H}_1(T)&=\bigg\langle\dfrac{\omega_q\bm{q}^2}{(\omega_q^2-\Delta_{p\Delta}^2+i\epsilon)^2}\bigg\rangle,\label{def:H1}\\
\mathcal{H}_2(T)&=\bigg\langle\dfrac{\omega_q\bm{q}^4}{(\omega_q^2-\Delta_{p\Delta}^2+i\epsilon)^2}\bigg\rangle,\label{def:H2}\\
\mathcal{G}_1(T)&=\Bigg\langle\dfrac{\bm{q}^2(\omega_q^2+\Delta_{p\Delta}^2)}{\omega_q(\omega_q^2-\Delta_{p\Delta}^2+i\epsilon)^2}\Bigg\rangle,\label{def:G1}\\    \mathcal{G}_2(T)&=\Bigg\langle\dfrac{\bm{q}^4(\omega_q^2+\Delta_{p\Delta}^2)}{\omega_q(\omega_q^2-\Delta_{p\Delta}^2+i\epsilon)^2}\Bigg\rangle,\label{def:G2}
\end{align}
where $\Delta_{p\Delta}$ is the mass difference between the proton and $\Delta^{++}$.
\begin{figure}
    \centering
    \includegraphics[width=1\linewidth]{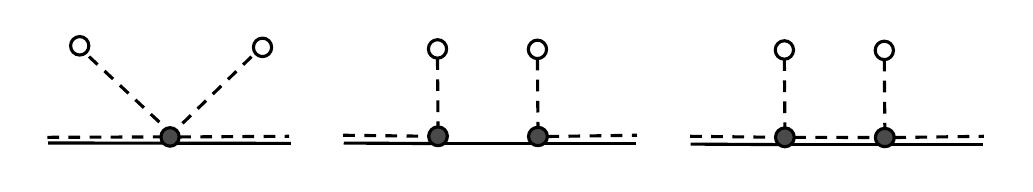}
    \caption{Diagrams of the $\Delta^{++}$ self energy from the pion forward scattering. $\Delta$, $p$, and $\pi$ are denoted by a double (solid+dashed) line, a solid line, and a dashed line, respectively. The small open circle represents the open of the loop, while the small solid circle denotes the vertex. The contribution to $i\Sigma^{ij}$ is the sum of the diagrams with $\Delta$ legs and pion legs amputated, weighted by $\bm{f}_\pi(\omega_q)/(2\omega_q)$, integrated over $\bm{q}$.}
    \label{fig:forward_scattering}
\end{figure}
\subsubsection{Self energy of $\Delta^{++}$ in the pion gas}
Fig.~\ref{fig:forward_scattering} presents the processes that contribute to the self energy of $\Delta^{++}$.  According to the finite temperature calculations in Ref.~\cite{Gao:2019idb}, the second diagram with two intermediate particles in Fig.~\ref{fig:forward_scattering} should practically include two thermal correction parts from the coherent forward scattering of an on-shell pion and an on-shell proton, respectively. However, the thermal part stemming from the coherent forward scattering of an on-shell proton with a large mass $M$ will produce a great exponential suppression factor, i.e., $e^{-M/T}$, which makes the thermal correction to the self energy tiny. Therefore, we ignore the thermal part resulting from the coherent forward scattering of an on-shell proton. The case in the third diagram is the same. Only the thermal part from the coherent forward scattering of an on-shell pion is considered in its corresponding self energy calculation. Therefore, Fig.~\ref{fig:forward_scattering} only shows that an on-shell pion is scattered from the open loop with momentum $\bm{q}$ and flavor $i$, and then it is scattered back into the other open loop with the same momentum $\bm{q}$ and flavor $i$ after it is modified by the pion gas. Hence, the self energy of each diagram must be weighted by a factor $\bm{f}_\pi(\omega_q)/(2\omega_q)$ to account for the effect of the pion gas before we sum them up, where $\bm{f}_\pi(\omega_q)$ is the pion momentum distribution and $1/(2\omega_q)$ is a normalization factor. In practical calculations, the thermal contributions of the pion gas can be obtained by applying a simple substitution to the pion propagator in the loop~\cite{Braaten:2023ciy}:
\begin{align}
    \frac{i}{q^2-m_{\pi }^2+i\epsilon}\to\bm{f}_\pi(|q^0|)2\pi\delta(q^2-m_\pi^2).
    \label{eq32:pion_line1}
\end{align}
The delta function can be expressed as
\begin{align}
\delta(q^2-m_\pi^2)=\sum_{\pm}\theta(\pm q^0)\frac{1}{2\omega_q}\delta(|q_0|-\omega_q), 
\label{eq32:pion_line2}
\end{align}
which is referred to as the cutting of the pion line. The cutting of the pion line
in the first diagram in Fig.~\ref{fig:forward_scattering} is 0, because the vertex in leading order is 0 when the incoming and outgoing
pions have the same flavor.
For the second and third diagrams in Fig.~\ref{fig:forward_scattering}, the corresponding vertices, which describe the interactions with an incoming pion, an incoming baryon (proton in the second diagram and $\Delta$ in the third diagram) and an outgoing $\Delta^{++}$, are given by 
\begin{align}
 &V_{p\pi^+\to\Delta^{++}}=\dfrac{g_{N\Delta\pi}}{F_\pi}q^\mu,\label{eq:vertex1}\\
 &V_{\Delta^{++}\pi^0\to\Delta^{++}}=\dfrac{g_1}{2F_\pi}\slashed{q}\gamma_5g^{\mu\nu},\label{eq:vertex2}\\
 &V_{\Delta^{+}\pi^+\to\Delta^{++}}=\dfrac{g_1}{\sqrt{2}F_\pi}\slashed{q}\gamma_5g^{\mu\nu},\label{eq:vertex3}
\end{align}
where $q$ is the four momentum of the pion, $F_\pi=92.4$~MeV is the pion decay constant, $g_{N\Delta\pi}$ is the coupling in the transition of $\Delta\to N\pi$, and $g_1$ is the coupling in the transition of $\Delta\to\Delta\pi$. The vertices of the corresponding inverse transitions with the outgoing pion can be obtained by replacing $q$ by $-q$. The relevant Lagrangians used to obtain the vertices in Eqs.~\eqref{eq:vertex1}, \eqref{eq:vertex2}, and \eqref{eq:vertex3} are shown in Appendix~\ref{Lagrangian}. 

With all preparations, one can obtain the self energy expressions for the last two diagrams in Fig.~\ref{fig:forward_scattering}. For the second diagram, its self energy has the form of
\begin{widetext}
\begin{align}
    \Sigma^{ij}_2(E,\bm{p},T)
    =&~\dfrac{g_{N\Delta\pi}^2\gamma^\alpha}{2m_pF_\pi^2}\Bigg\langle \dfrac{1}{2\omega_q}\sum_{\pm}\dfrac{q^\alpha q^i q^j}{E-\varepsilon\pm \omega_q+\Delta_{p\Delta}-(\bm p\pm\bm q)^2/(2m_p)+i\epsilon}\Bigg\rangle\notag\\
    &-\dfrac{g_{N\Delta\pi}^2I}{F_\pi^2}\Bigg\langle \dfrac{1}{2\omega_q}\sum_{\pm}\dfrac{ q^i q^j}{E-\varepsilon\pm \omega_q+\Delta_{p\Delta}-(\bm p\pm\bm q)^2/(2m_p)+i\epsilon}\Bigg\rangle,
    \label{eq:sigma_2}
\end{align}
\end{widetext}
where $m_p$ is the mass of the proton and $\Delta_{p\Delta}=M-m_p$. It is noted that we employ the nonrelativistic form of the proton propagator 
\begin{align}
    iS^{\text{NR}}_F(k)=\dfrac{iI}{k^0-m_p-\bm{k}^2/(2m_p)+i\epsilon}
\end{align}
in the calculation of Eq.~\eqref{eq:sigma_2}. The terms $\omega_q$ and $\Delta_{p\Delta}$ in the denominator of the self energy in Eq.~\eqref{eq:sigma_2} are of the order $m_\pi$, while the small energies $E-\varepsilon$ and $(\bm{p}\pm\bm{q})^2/(2m_p)$ are considered to be of the order $m_\pi^2/m_p$. Therefore, we can perform a heavy meson expansion by expanding Eq.~\eqref{eq:sigma_2} in powers of the above two small energies. Terms with odd powers of $\bm{q}$ vanish after averaging over the directions of the pion momentum. Additionally, one can notice that $iS_0^{ij}$ in Eq.~\eqref{propagator_Delta_NR} only has two diagonal elements in the upper left block, which means that only the two diagonal elements in the upper left block of $i\Sigma^{ij}$ contribute to the correction of the pole of $iS^{ij}$ by referring to Eq.~\eqref{eq:S_ij1}. 

Hence, the leading order term of $\Sigma_2^{ij}(E,\bm{p},T)$ contributing to the pole correction is independent of $E$ and $\bm{p}$:
\begin{align}
    \Sigma_2^{\text{LO}}(T)=\dfrac{g_{N\Delta\pi}^2\bm{n}_\pi}{6mF_\pi^2}\mathcal{I}^*(T)-\dfrac{g_{N\Delta\pi}^2\Delta_{p\Delta}\bm{n}_\pi}{3F_\pi^2}\mathcal{F}^*(T),
    \label{eq:Sigma2_LO}
\end{align}
where $\mathcal{H}^*(T)$ and $\mathcal{F}^*(T)$ are the complex conjugates of the thermal averages of $\mathcal{H}(T)$ and $\mathcal{F}(T)$ defined in Eq.~\eqref{def:I} and Eq.~\eqref{def:F}, respectively. And their imaginary parts
\begin{align}
    \text{Im}[\mathcal{I}(T)]&=\dfrac{\bm{f}_\pi(\Delta_{p\Delta})}{4\pi\bm{n}_\pi}\Big(-\Delta_{p\Delta} 
   q_c^3\Big)\theta(\Delta_{p\Delta}-m_\pi),\label{eq:ImI}\\
    \text{Im}[\mathcal{F}(T)]&=\dfrac{\bm{f}_\pi(\Delta_{p\Delta})}{4\pi\bm{n}_\pi}\Big(-\dfrac{1}{\Delta_{p\Delta}}q_c^3\Big)\theta(\Delta_{p\Delta}-m_\pi),\label{eq:ImF}
\end{align}
with $q_c=\sqrt{\Delta_{p\Delta}^2-m_\pi^2}$ are calculated to obtain $\text{Im}[\delta\varepsilon]$. Here $\theta(x)$ is the step function. After removing the terms which do not affect the pole of the $\Delta^{++}$ propagator, 
the next-to-leading order term of $\Sigma_2^{\text{NLO}}$ is a linear function of $E$ and $\bm{p}^2$:
\begin{widetext}
 \begin{align}
\Sigma_2^{\text{NLO}}(E,\bm{p},T)=&-\dfrac{g_{N\Delta\pi}^2\Delta_{p\Delta}\bm{n}_\pi}{3m_pF_\pi^2}\Big(\dfrac{\bm{p}^2}{2m_p}-E+\varepsilon\Big)\mathcal{H}_1^*(T)-\dfrac{g_{N\Delta\pi}^2\Delta_{p\Delta}\bm{n}_\pi}{6m_p^2F_\pi^2}\mathcal{H}_2^*(T)\notag\\
&+\dfrac{g_{N\Delta\pi}^2\bm{n}_\pi}{3F_\pi^2}\Big(\dfrac{\bm{p}^2}{2m_p}-E+\varepsilon\Big)\mathcal{G}_1^*(T)+\dfrac{g_{N\Delta\pi}^2\bm{n}_\pi}{6m_pF_\pi^2}\mathcal{G}_2^*(T),
\label{eq:Sigma2_NLO}
\end{align}   
\end{widetext}
where $\mathcal{H}_1(T)$, $\mathcal{H}_2(T)$, $\mathcal{G}_1(T)$, and $\mathcal{G}_2(T)$ are the complex conjugates of the thermal averages defined in Eqs.~\eqref{def:H1}, ~\eqref{def:H2}, ~\eqref{def:G1}, and ~\eqref{def:G2}. And their imaginary parts are defined as
\begin{align}
    \text{Im}[\mathcal{H}_1(T)]&=-\dfrac{1}{8\pi \bm{n}_\pi}\Bigg(\dfrac{f_\pi(\Delta_{p\Delta})q_c^3}{\Delta_{p\Delta}}+3f_\pi(\Delta_{p\Delta})\Delta_{p\Delta} q_c\Bigg)\notag\\
    &~~~\times\theta(\Delta_{p\Delta}-m_\pi),\label{eq:ImH1}\\
    \text{Im}[\mathcal{H}_2(T)]&=-\dfrac{1}{8\pi \bm{n}_\pi}\Bigg(\dfrac{f_\pi(\Delta_{p\Delta})q_c^5}{\Delta_{p\Delta}}+5f_\pi(\Delta_{p\Delta})\Delta_{p\Delta} q_c^3\Bigg)\notag\\
    &~~~\times\theta(\Delta_{p\Delta}-m_\pi),\label{eq:ImH2}\\
    \text{Im}[\mathcal{G}_1(T)]&=-\dfrac{1}{4\pi\bm{n}_\pi}3q_c\bm{f}_\pi(\Delta_{p\Delta})\Delta_{p\Delta}\theta(\Delta_{p\Delta}-m_\pi),\label{eq:ImG1}\\
  \text{Im}[\mathcal{G}_2(T)]&=-\dfrac{1}{4\pi\bm{n}_\pi}5q_c^3\bm{f}_\pi(\Delta_{p\Delta})\Delta_{p\Delta}\theta(\Delta_{p\Delta}-m_\pi).\label{eq:ImG2}
\end{align}
The detailed calculation of the imaginary parts of the thermal averages can be found in Appendix ~\ref{calculation}. 

Following the same procedure of the self energy calculation for the second diagram, one can also obtain the self energy expression with thermal correction from the pion gas for the third diagram, but it does not modify $\text{Im}[\delta\varepsilon]$, since the internal $\Delta$ particles ($\Delta^+$ or $\Delta^{++}$) cannot be on-shell with the pions in the loop simultaneously. Therefore, we only need to consider the correction to the pole of $\Delta^{++}$ propagator from $\Sigma_2(E,\bm{p},T)$.

\subsubsection{The extraction of thermal width shift of $\Delta^{++}$}
The denominator of Eq.~\eqref{eq:S_ij2} can be rewritten as
 \begin{align}
    &E-\varepsilon-\dfrac{\bm{p}^2}{2M_\Delta}-\Sigma_2(E,\bm{p},T)\notag\\
    =&(1+\delta Z)^{-1}\bigg[E-\varepsilon-\delta\varepsilon-(1+\zeta)\dfrac{\bm{p}^2}{2M_\Delta}+\ldots\bigg]\label{eq:comparison}
\end{align}   
with 
\begin{align}
   \Sigma_2(E,\bm{p},T)=\Sigma_2(T)^{\text{LO}}+\Sigma_2(E,\bm{p},T)^{\text{NLO}}. 
\end{align}
The correction $\delta\varepsilon$ can be obtained from the expansion in Eq.~\eqref{eq:comparison}:
\begin{align}
    \delta\varepsilon(T)=&~\dfrac{g_{N\Delta\pi}^2\bm{n}_\pi}{6m_pF_\pi^2}\mathcal{I}^*(T)-\dfrac{g_{N\Delta\pi}^2\Delta_{p\Delta}\bm{n}_\pi}{3F_\pi^2}\mathcal{F}^*(T)\notag\\
    &-\dfrac{g_{N\Delta\pi}^2\Delta_{p\Delta}\bm{n}_\pi}{6m_p^2F_\pi^2}\mathcal{H}_2^*(T)+\dfrac{g_{N\Delta\pi}^2\bm{n}_\pi}{6m_pF_\pi^2}\mathcal{G}_2^*(T).\label{eq:delta_varepsilon}
\end{align}
By replacing $\mathcal{I}^*(T)$, $\mathcal{F}^*(T)$, $\mathcal{H}_2^*(T)$, and $\mathcal{G}_2^*(T)$ in Eq.~\eqref{eq:delta_varepsilon} by the complex conjugates of Eqs.~\eqref{eq:ImI}, \eqref{eq:ImF}, \eqref{eq:ImH2}, and \eqref{eq:ImG2}, one can obtain $\text{Im}[\delta\varepsilon(T)]$. Therefore, The thermal width shift of $\Delta^{++}$ is represented as 
\begin{align}
  \delta\Gamma(T)=-2\text{Im}[\delta\varepsilon(T)] . 
\end{align} The values of the parameters 
\begin{align}
    &m_\pi=138.04~\text{MeV},~~~m_p=938.27~\text{MeV},\notag\\
    &M_{\Delta}=1230.55~\text{MeV}
\end{align}
used in the calculations are from PDG~\cite{ParticleDataGroup:2020ssz}.

\section{Results and Discussions}\label{Results and Analysis} 
\begin{figure}
    \centering
    \includegraphics[width=0.9\linewidth]{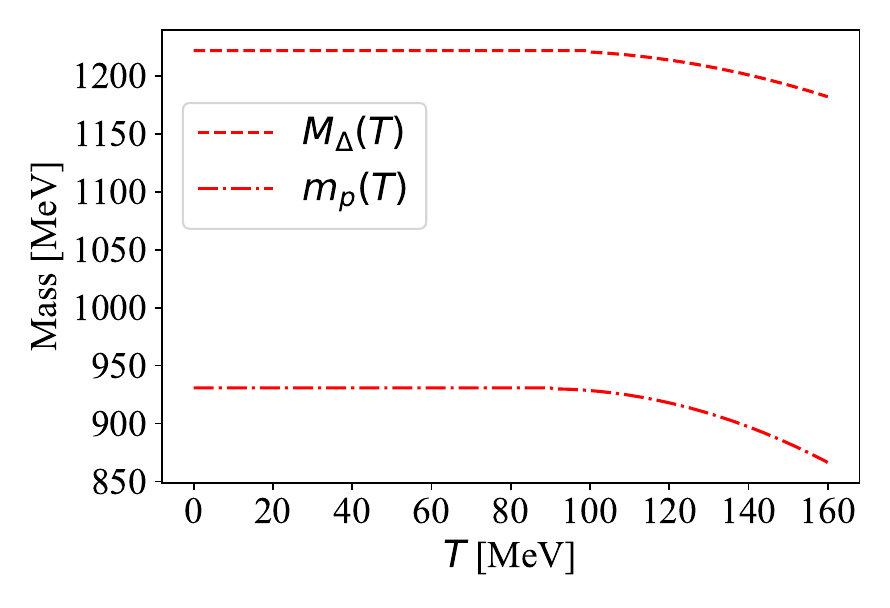}
    \caption{The thermal shift of $\Delta$ (dashed line) and $p$ (dot-dashed line) masses extracted from Ref.~\cite{Torres-Rincon:2015rma}.}
    \label{fig:Massft}
\end{figure}
\begin{figure}
    \centering
    \includegraphics[width=0.9\linewidth]{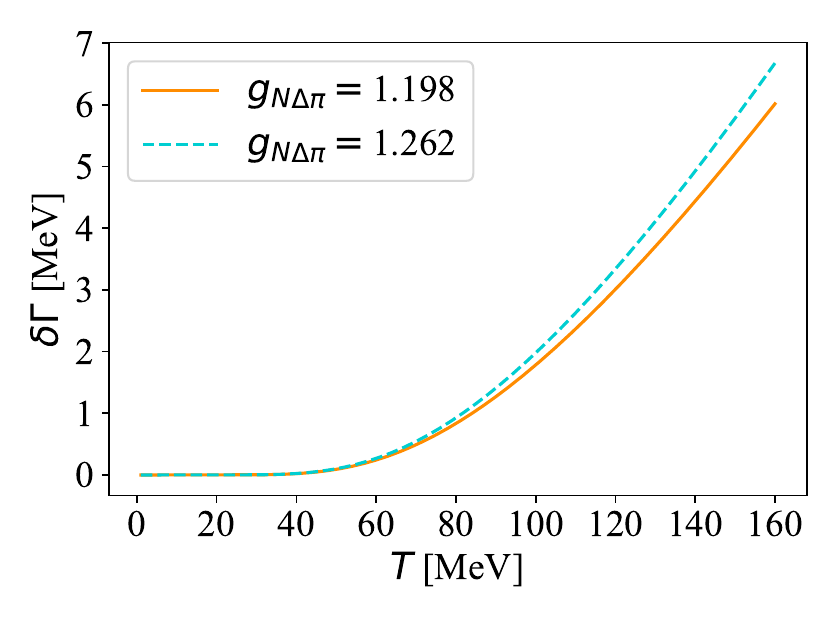}
    \includegraphics[width=0.9\linewidth]{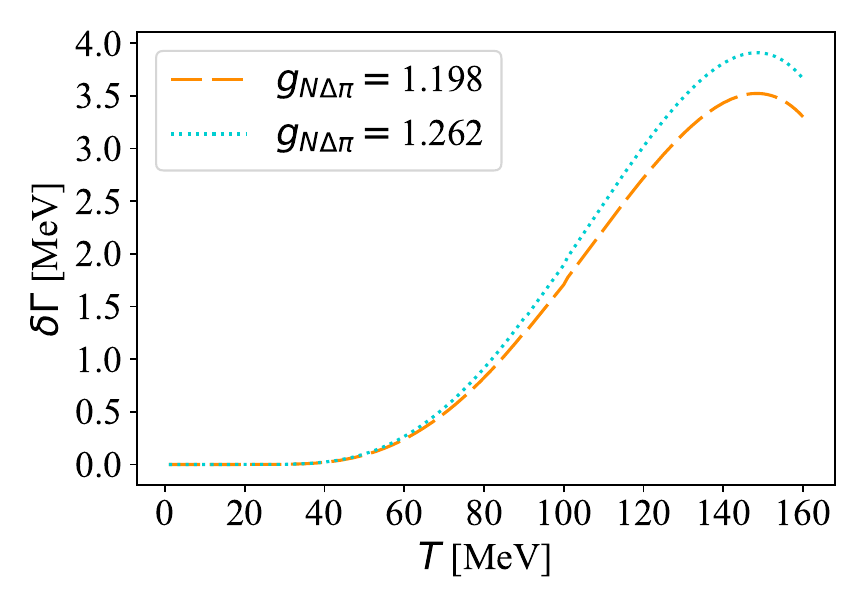}
    \caption{The thermal width shift of $\Delta^{++}$ without (upper panel) and with (lower panel) the thermal corrections to $m_p$ and $M_\Delta$ extracted from Ref.~\cite{Torres-Rincon:2015rma}. The orange and blue lines are the thermal width shifts  obtained by the lower and upper limits of the coupling $g_{N\Delta\pi}$, respectively.}
    \label{fig:Gamma&T}
\end{figure}

To extract the thermal width shift of $\Delta^{++}$ baryon, the coupling $g_{N\Delta\pi}$ should be provided. 
Its lower and upper limits, i.e., $g_{N\pi\Delta}=1.198$ and $g_{N\pi\Delta}=1.262$, can be extracted from the lower and upper limits of physical partial width of $\Delta\to N\pi$~\cite{Chen:2025bgq}. 
With those two values, the thermal width shift as a function of temperature $T$ is illustrated in Fig.~\ref{fig:Gamma&T}, where the orange solid and blue dashed lines correspond the lower and upper limits, respectively.
It is obvious that the width of $\Delta^{++}$ would increase as the temperature increases, indicating the dissociation of $\Delta^{++}$ at a sufficiently high temperature.  
 The behavior of this thermal width property is in agreement with those of other particles at finite temperature investigated using different theoretical methods, such as pion and nucleon studied with the correlation function of a three-quark current~\cite{Leutwyler:1990uq}, charmed mesons and bottomed mesons studied in lattice QCD~\cite{Torres-Rincon:2026rso}.

As the temperature of hadron phase is as high as about $160~
\mathrm{MeV}$, one can see that the width shift due to the surrounding pions can be as large as $6~
\mathrm{MeV}$, as shown in the upper panel of Fig.~\ref{fig:Gamma&T}. This value is a key value for any transport model to predict correct pion and nucleon yields. Assuming the phase-space distributions of $\Delta$, nucleon and pion are stable with the temperature, the increasing width of $\Delta$ means the large $N\pi\to\Delta$ cross section due to the equilibrium equation, i.e., Eq.~\eqref{eq:equilibrium}, as we expect. 

There is an argument that the masses of octet and decuplet baryons will become degenerate when chiral symmetry is restored at sufficient high temperature~\cite{Cohen:1989qe, Brown:2001nh}, making the phase space of the $\Delta\to N\pi$ smaller. To investigate this effect, we borrow the temperature dependence of the $\Delta$, nucleon and pion mass in Ref.~\cite{Torres-Rincon:2015rma} based on the NJL model. As shown by Fig.~\ref{fig:Massft}, the masses of $\Delta$ and nucleon are stable with the increasing of temperature below $100~\mathrm{MeV}$. In addition, the pion mass is stable below $230~\mathrm{MeV}$~\cite{Torres-Rincon:2015rma}. In this case, the behavior of the width shift of $\Delta$ baryon, the lower panel of Fig.~\ref{fig:Gamma&T}, is the same as that in the upper panel of Fig.~\ref{fig:Gamma&T}. 
When the temperature further increases above $100~\mathrm{MeV}$, the masses of $\Delta$ and nucleon decrease, making the width shift different from that in the upper panel. Especially, when the temperature is around $140~\mathrm{MeV}$, the width shift arrives its  maximum value due to the non-linear behavior of the two-body phase space.  

When comparing with the experimental data from STAR Collaboration, i.e., the width of $\Delta^{++}$ at $|y|<0.5$ ($y$ is the rapidity) for minimum bias $d$+Au interactions as a function of the horizontal momentum $p_t$ \cite{STAR:2008twt}, $p_t$ can be approximated as the three momentum $|\bm{p}|$ due to $|y|<0.5$.
Since the thermal width shift of $\Delta^{++}$, i.e., $\delta\Gamma(T)$, is obtained in the nonrelativistic framework, we only take the experiment data with $p_t<1$~GeV. Table \ref{tab:pt&gamma} lists the data we used in the calculation. The $\Delta^{++}$ width in vacuum is taken as  $-2$ times the imaginary part of its pole position, i.e., $\Gamma_0=97.37$ MeV~\cite{ParticleDataGroup:2020ssz}, so the width with the thermal correction from the pion gas is given by
\begin{align}
    \Gamma(T)=\Gamma_0+\delta\Gamma(T).\label{eq:width_ft}
\end{align}
Substituting the width shown in the second column in Table~\ref{tab:pt&gamma} into Eq.~\eqref{eq:width_ft}, we can obtain the corresponding temperature $T$. In practical calculations, we also consider the statistical error of the experimental width shown in the third column of Table \ref{tab:pt&gamma}.
\begin{table}[h]
\renewcommand{\arraystretch}{1.2}
    \centering
    \begin{tabular}{cccc}
    \hline\hline
    $p_t$~[GeV]&$\Gamma$~[GeV]&Statistical error~[GeV]&$\delta\Gamma$~[MeV]\\
    \hline
         0.500&  0.099&   0.007&1.630\\
         0.700&  0.110&   0.005&12.630\\
         0.900&  0.117&   0.004&17.630\\
    \hline
    \end{tabular}
    \caption{The width of $\Delta^{++}$ as a function of $p_t$ at $|y|<0.5$ for minimum bias $d$+Au collisions. The data  are taken from Ref.~\cite{STAR:2008twt}.}
    \label{tab:pt&gamma}
\end{table}
If the observed width shift were attributed entirely to the surrounding pion gas, reproducing a shift of $1.630-17.630$~MeV (as listed in Table~\ref{tab:pt&gamma})  would demand a temperature of 97~MeV-306~MeV for $g_{ N\Delta\pi}=1.198$---far exceeding the hadronic-phase temperature. Moreover, part of this apparently high temperature can be attributed to collective flow; That is the slope parameter of the $m_T$-spectra is not a pure thermal temperature but receives an additive flow contribution~\cite{Schnedermann:1993ws}. This suggests that additional mechanisms beyond the pion gas must contribute to the in-medium $\Delta$ width shift.

The STAR Collaboration also investigates the width variation of $\Lambda^*$ with $J^P=\dfrac{3}{2}^-$ as a function of $p_t$ ~\cite{STAR:2008twt}. For comparison, we consider the modification from the $\Lambda^*\to\Sigma\pi$ transition to the self energy of $\Lambda^*$ in the pion gas and obtain its corresponding thermal width correction. However, the thermal width shift is slight and cannot account for the large increase in width of $\Lambda^*$ with increasing $p_t$. Practically, there is another dominant transition of $\Lambda^*$, i.e., $\Lambda^*\to N\bar{K}$, which will also greatly contribute to the pole position of $\Lambda^*$ compared to the $\Lambda^*\to\Sigma\pi$ transition. Thus, considering the correction from the kaon gas may compensate for the large difference between the width in vacuum and at finite temperature. 

\section{Summary}\label{summary}
We compute the thermal width shift of the $\Delta^{++}$ resonance induced by a pion gas within a nonrelativistic effective field theory framework. The self-energy of $\Delta^{++}$ is evaluated from pion-forward-scattering diagrams with intermediate proton and $\Delta$ states, weighted by the thermal pion distribution. Analytical expressions for the imaginary part of the self-energy are obtained, yielding the temperature-dependent width correction $\delta\Gamma(T)$.

The $\Delta^{++}$ width increases with temperature, reaching a shift of approximately $6$~MeV at $T \approx 160$~MeV, consistent with the expected thermal broadening of hadronic resonances. When the temperature dependence of the $\Delta$ and proton masses from an NJL-model chiral restoration scenario is included, the width shift develops a non-monotonic behavior with a maximum near $T \approx 140$~MeV, reflecting the competition between collisional broadening and the shrinking $\Delta \to N\pi$ phase space as the $N$--$\Delta$ mass gap closes.

We apply our formalism to STAR data for $\Delta^{++}$ in d+Au collisions at $\sqrt{s_{NN}} = 200$~GeV. The temperature extracted at $p_t = 900$~MeV is $T \approx 300$~MeV, significantly exceeding the hadronic-phase temperature. This explicitly demonstrates that the experimentally observed width shift is not solely driven by the surrounding pion gas---additional contributions from genuine hot-medium effects and collective radial flow must play an important role.

By isolating the pion-gas contribution to the resonance width shift, our framework provides a potential method for temperature extraction in heavy-ion collisions that complements traditional approaches based on spectral slopes and particle ratios. The doubly charged $\Delta^{++}$ channel offers particular experimental advantages due to its clean $p + \pi^+$ decay signature. The remaining width shift, after subtracting the pion-gas contribution computed here, can serve as a cleaner proxy for the genuine QGP and flow effects.

\section*{Acknowledgments}
We are grateful to Deliang Yao, Quanxing Ye, and Zaochen Ye for the helpful discussion. 
This work is partly supported by the National Natural Science Foundation of China with Grants No.~12375073, No.~12547105, No.~12505110, and No.~12422509.

\appendix
\section{The fields and propagators}\label{fields}
$\Psi_\mu^i$ is a shorthand notation for $\Psi_{\mu,\alpha,i,r}$, which is a vector-spinor isovector-isospinor field, with $\mu$ being a Lorentz vector, $\alpha$ Dirac spinor index, $i$ an isovector index, and $r$ an isospinor index. From now on, the Dirac spinor and the isospinor indices will be suppressed for simplicity. The fields $\Psi_\mu^i$ are related to the physical $\Delta(1232)$ states $\Delta^{++}$, $\Delta^+$, $\Delta^0$ and $\Delta^-$ by~\cite{Yao:2016vbz} 
\begin{align}
    \xi_{1j}^{3/2}\Psi_\mu^j&=\frac{1}{\sqrt{2}}\begin{bmatrix}
        \frac{1}{\sqrt{3}}\Delta^0-\Delta^{++}\\
        \Delta^{-}-\frac{1}{\sqrt{3}}\Delta^+
    \end{bmatrix}_\mu,\notag\\
    \xi_{2j}^{3/2}\Psi_\mu^j&=-\frac{i}{\sqrt{2}}\begin{bmatrix}
        \frac{1}{\sqrt{3}}\Delta^0+\Delta^{++}\\
        \Delta^{-}+\frac{1}{\sqrt{3}}\Delta^+
    \end{bmatrix}_\mu,\notag\\
    \xi_{3j}^{3/2}\Psi_\mu^j&=\sqrt{\frac{2}{3}}\begin{bmatrix}
        \Delta^{+}\\
        \Delta^0
    \end{bmatrix}_\mu,
\end{align}
where the isospin-$\frac{3}{2}$ projection operator is defined as $\xi_{ij}^{3/2}=\delta_{ij}-\tau_i\tau_j/3$.
The propagator for the $\Delta$ particle is expressed as
\begin{align}
    iS_{0,ij}^{\mu\nu}(p)=&-\frac{i(\slashed{p}+m_\Delta)}{p^2-m_\Delta^2+i\epsilon}\Big[g^{\mu\nu}-\frac{1}{d-1}\gamma^\mu\gamma^\nu
    +\frac{1}{(d-1)m_\Delta}\notag\\&\times(p^\mu\gamma^\nu-\gamma^\mu p^\nu)-\frac{d-2}{(d-1)m_\Delta^2}p^\mu p^\nu\Big]\xi^{3/2}_{ij}.
\end{align}

$\Psi_N$ represents the nucleons $p$ and $n$, and has the form of 
\begin{align}
  \Psi_N=\begin{bmatrix}
        p\\
        n
    \end{bmatrix}. 
\end{align}
The expression of nucleon propagator is given by
\begin{align}
    iS_F(k)=\frac{i(\slashed{k}+m_N)}{k^2-m_N^2+i\epsilon}.
\end{align}

The pion propagator is written as 
\begin{align}
    i\Delta^{ab}_F(q)=\frac{i\delta^{ab}}{q^2-m_\pi^2+i\epsilon},
\end{align}
where $a$ and $b$ are isospin indices for pions in the loop.

\section{Lagrangian for the interaction vertices of $\pi\Delta\to\pi\Delta$, $\pi N\to\Delta$, and $\pi\Delta\to\Delta$ transitions}\label{Lagrangian}
The Lagrangian for $\pi\Delta\to\pi\Delta$ transition in leading order: 
\begin{align}
  \mathcal{L}_{\pi\Delta\to\pi\Delta}=&-\bar{\Psi}^i_\mu\xi^{3/2}_{ij}\big\{i\slashed{D}^{jk}g^{\mu\nu}-i\big(\gamma^\mu D^{\nu,jk}+\gamma^\nu D^{\mu,jk}\big)\notag\\
  &+i\gamma^\mu\slashed{D}^{jk}\gamma^\nu \big\}\xi^{3/2}_{kl}\Psi^l_\nu,
  \label{eq:pi_Delta_pi_Delta}
\end{align}
where 
\begin{align}
    D_{\mu,ij}&=-\frac{1}{2F_\pi^2}\epsilon_{ijk}\epsilon_{lmk}(\partial_\mu\phi^l)\phi^m+\delta_{ij}\frac{i}{4F_\pi^2}\epsilon_{abc}\phi^a(\partial_\mu\phi^b)\tau^c,\\
    \slashed{D}_{ij}&= D_{\mu,ij}\gamma^\mu.
\end{align}

The Lagrangian for $\pi N\to\Delta$ transition in lading order:
\begin{align}
    \mathcal{L}_{\pi N\Delta}=h\bar{\Psi}_\mu^i\xi_{ij}^{3/2}\Theta^{\mu\alpha}(z_1)\omega_\alpha^j\Psi_N+h\bar{\Psi}_N\omega_\alpha^{j\dagger}\Theta^{\alpha\mu}(z_1)\xi_{ji}^{3/2}\Psi_\mu^i.
    \label{eq:pi_N_Delta}
\end{align}
where $\Theta^{\mu\alpha}(z_1)=g^{\mu\alpha}+z_1\gamma^\mu\gamma^\alpha$ with $z_1=0$. $h$ is the bare pion-nucleon-delta coupling constant at lowest order, and will be replaced by the physical value $g_{N\Delta\pi}$. For only one pion field, $
 \omega_\alpha^i=\frac{1}{2}\langle\tau^iu_\alpha\rangle\simeq-\frac{1}{F}\partial_\alpha\pi^i$.

The Lagrangian for the $\pi\Delta\to\Delta$ transition in leading order is 
\begin{align}
    \mathcal{L}^{(1)}_{\pi\Delta\Delta}
    =\frac{g_1}{2F_\pi}\bar{\Psi}^i_\mu\xi^{3/2}_{ij}\slashed{\phi}\delta^{jk}\gamma_5g^{\mu\nu}\xi^{3/2}_{kl}\Psi^l_\nu.
    \label{eq:pi_Delta_Delta}
\end{align}

\section{The calculation of the imaginary part of the thermal average}\label{calculation}
Some of the thermal averages over the pion momentum involve integrals of the form
\begin{align}
    \mathcal{F}_{n}(\sigma)=\lim_{\epsilon\to0^+}\int_0^\infty \text{d}qF(\bm{q}^2)\frac{1}{(\bm{q}^2-\sigma+i\epsilon)^n},
    \label{fn}
\end{align}
where $F(\bm{q}^2)$ is a real-valued function that is smooth as $\bm{q}^2\to0$ and decreases rapidly to 0 as $\bm{q}^2\to\infty$. The real parameter $\sigma$, which can be positive or negative.

In the case $n=1$, Eq.~\eqref{fn} can be expressed as the sum of a principle-value integral and the integral of a delta function if $\sigma>0$:
\begin{align}
     \mathcal{F}_{1}(\sigma)=&\lim_{\epsilon\to0^+}\int_0^\infty \text{d}|\bm{q}|F(\bm{q}^2)\frac{1}{\bm{q}^2-\sigma+i\epsilon}\notag\\
     =&\int_0^\infty \text{d}|\bm{q}|F(\bm{q}^2)\left(\mathcal{P}\frac{1}{\bm{q}^2-\sigma}-i\pi\delta(\bm{q}^2-\sigma)\right)\notag\\
     =&\int_0^\infty \text{d}|\bm{q}|\frac{F(\bm{q}^2)-F(\sigma)}{\bm{q}^2-\sigma}- i\frac{\pi}{2\sqrt{\sigma}}F(\sigma)\theta(\sigma),
     \label{f1s}
\end{align}
where we have exploited the following relations
\begin{align}
    \delta(f(x)-f(x_0))&=\delta(x-x_0)/|f'(x_0)|,\\
    \mathcal{P}\int_0^\infty\frac{\text{d}|\bm{q}|f(|\bm{q}|)}{\bm{q}^2-\bm{k}^2}&=\int_0^\infty\frac{f(|\bm{q}|)-f(|\bm{k}|)}{\bm{q}^2-\bm{k}^2}\text{d}|\bm{q}|
\end{align}
with $\mathcal{P}$ denoting the principle-value integral. Therefore, the imaginary part of $\mathcal{F}_{1}(\sigma)$ can be obtained by
\begin{align}
    \text{Im}[\mathcal{F}_{1}(\sigma)]=-\dfrac{\pi}{2\sqrt{\sigma}}F(\sigma)\theta(\sigma).
\end{align}

The case of $n=2$ of Eq.~\eqref{fn} can be reduce to the case $n=1$ by integrating by parts:
\begin{align}
    \mathcal{F}_{2}(\sigma)=&\lim_{\epsilon\to0^+}\int_0^\infty \text{d}|\bm{q}|F(\bm{q}^2)\frac{1}{(\bm{q}^2-\sigma+i\epsilon)^2}\notag\\
    =&-\lim_{\epsilon\to0^+}\int_0^\infty \text{d}|\bm{q}|F(\bm{q}^2)\frac{\text{d}}{\text{d}\bm{q}^2}\frac{1}{\bm{q}^2-\sigma+i\epsilon}\notag\\
    =&\int_0^\infty \text{d}|\bm{q}|\left[F'(\bm{q}^2)-\frac{F(\bm{q}^2)}{2\bm{q}^2}\right]\left(\mathcal{P}\frac{1}{\bm{q}^2-\sigma}-i\pi\delta(\bm{q}^2-\sigma)\right)\notag\\
    =&\int_0^\infty \text{d}|\bm{q}|\left[F'(\bm{q}^2)-\frac{F(\bm{q}^2)}{2\bm{q}^2}-\left(F'(\sigma)-\frac{F(\sigma)}{2\sigma}\right)\right]\frac{1}{\bm{q}^2-\sigma}\notag\\
    &-i\frac{\pi}{2\sqrt{\sigma}}\left(F'(\sigma)-\frac{F(\sigma)}{2\sigma}\right)\theta(\sigma).
    \label{f2s}
\end{align}
So the imaginary part of $\mathcal{F}_{2}(\sigma)$ is given by
\begin{align}
    \text{Im}[\mathcal{F}_{2}(\sigma)]=-\dfrac{\pi}{2\sqrt{\sigma}}\left(F'(\sigma)-\frac{F(\sigma)}{2\sigma}\right)\theta(\sigma).
\end{align}
\nocite{*}

\end{document}